\documentclass[prc,aps,twocolumn,tightenlines,showpacs,nofootinbib,preprintnumbers,
amsmath,amssymb]{revtex4-1}

\usepackage[usenames,dvipsnames]{color}

\usepackage{graphicx}
\usepackage{bm}

\begin{document}

\title{Characteristics of a thermal neutrons scintillation detector with \\
the [ZnS(Ag)+$^6$LiF] at different conditions of measurements}

\author{
V.V.~Alekseenko$^{a}$,
I.R.~Barabanov$^{a}$,
R.A.~Etezov$^{a}$,
Yu.M.~Gavrilyuk$^{a}$, \\
A.M.~Gangapshev$^{a}$, 
A.M.~Gezhaev$^{a}$,
V.V.~Kazalov$^{a}$,
A.Kh.~Khokonov$^{b}$, \\
V.V.~Kuzminov$^{a}$, 
S.I.~Panasenko$^{c}$,
S.S.~Ratkevich$^{c}$
}
\affiliation{\small $^a$ Institute for Nuclear Research, RAS, Russia \\
$^b$ Kh.M.~Berbekov Kabardino-Balkarian State University, Russia \\
\small $^c$ V.N.~Karazin Kharkiv National University, Ukraine}


\begin{abstract}

A construction of a thermal neutron testing detector with a thin [ZnS(Ag)+$^6$LiF] scintillator is described.
Results of an investigation of sources of the detector pulse origin and the pulse features in a ground and underground conditions are presented.
Measurements of the scintillator own background, registration efficiency and a neutron flux at different objects of the BNO INR RAS were performed.
The results are compared with the ones measured by the $^3$He proportional counter.

\end{abstract}


\maketitle


\section{\label{intro}Introduction}
Large area scintillation detector of the thermal neutrons on a base of a thin [ZnS(Ag)+ $^6$LiF] scintillator has been put into operation at the Baksan Neutrino Observatory of the INR RAS at the last time \cite{p1,p2}. The neutron registration is occurred as a result of $^6$Li($n,\alpha$)$^3$H+4786 keV reaction. The cross section is 945 b \cite{p3}. A kinetic energy of the reaction products ($E_\alpha=2051$ keV, $E_{\rm H}=2735$ keV) converted at a light flash by the scintillator.
The scintillator components enter into the composition as an alloy.
Spectrometric characteristics of such detectors are known not enough.
The aim of the work was to investigate and refinement these parameters.

\section{Detector construction}

\begin{figure*}[pt]
\includegraphics*[width=4.25 in,angle=0.]{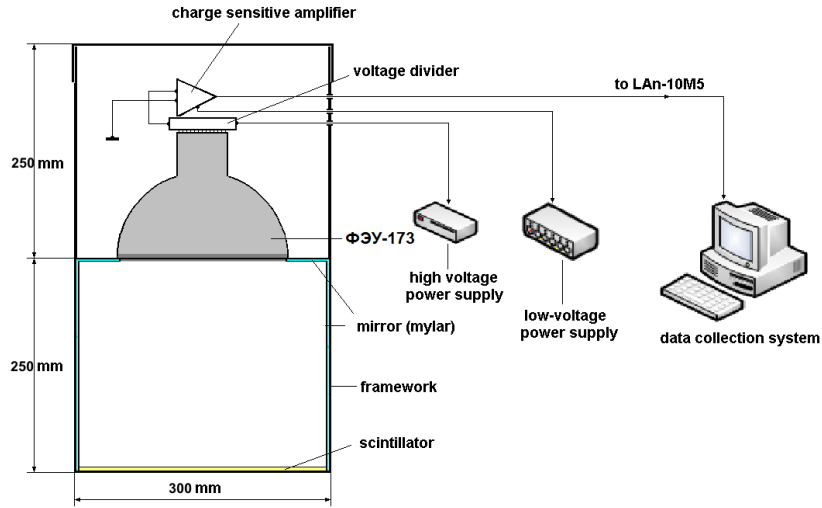}%
\caption{\label{fig:schematic_view} A schematic view of the test detector with the [ZnS(Ag)+$^6$LiF] scintillator and its electronics.}
\end{figure*}

A schematic view of the test detector and its electronics are sown on the Fig.~\ref{fig:schematic_view}.
The detector is assembled in a rectangular case with $30\times30\times50$ cm$^3$ sizes
made from  galvanized iron  of  the 0.7 mm thickness.
The covers of the top  and bottom are detachable.
A dividing plate with a central 150 mm diameter hole aimed to install photomultiplier FEU-173 (PMT) mounted in the middle of the case.
A charge sensitive preamplifier (CSP) with $\sim100$ $\mu$s self-discharge time is installed on the wall inside of the upper part of the case.
Pulses from the PMT’s anode resistor (4.8 M$\Omega$) go to the CSP input and further at the input of the LAn-10M5 digital oscilloscope card and are recorded into the personal computer (PC) memory.
A sampling frequency was equal to 6.25 MHz. 
The flat flexible plate with the [ZnS(Ag)+$^6$LiF] scintillator is placed on the floor of the bottom section.
It contains of the white sheet of plastic film ($207\times295$ mm$^2$ ) with a sticky side covered by grains of the scintillator with an average thickness of $\sim0.1$ mm \cite{p2}.
The sheet is laminated by the lavsan films. A surface of the bottom section covered with a reflecting mylar film for a better light collection. Densities of the pure ZnS(Ag) and $^6$LiF are equal to 4.09 g$\cdot$cm$^{-3}$ \cite{p4} and $\sim2.64$ g$\cdot$cm$^{-3}$ \cite{p7}.
A density of the mixture in the proportion $1:3$ is $\sim3$ g$\cdot$cm$^{-3}$.

\section{Results of measurements}

\begin{figure}[pt]
\includegraphics*[width=2.1in,angle=0.]{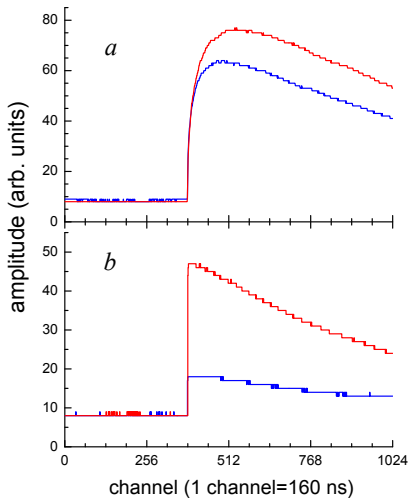}%
\caption{\label{fig:pulses}
The two types of pulses with test the detector. ``\emph{a}'' - the long pulse rise time, ``\emph{b}'' - the short pulse rise time.
}
\end{figure}

The two types of working pulses shown on the Fig.~\ref{fig:pulses} ($a,b$) were observed in a measurement with the detector at the ground laboratory. The first type ($a$) has a front time duration $\tau_f = 16-25$~$\mu$s which  corresponds to the own de-excitation time of the fine-grained scintillator $\tau_s= 8-10$~$\mu$s ($\tau_f = 3 \tau_s$). The CSP integrates an input current and a pulse maximum is shapes at the point where the current charging velocity turns equal to the preamplifier own charge decay velocity.
The second type (\emph{b}) of a pulse has a front time duration $\tau_f = 0.8 \tau_s$.
Its shape is similar to the noise pulses which occur after an irradiation of the scintillator and  PMT by an external light during of an adjustment work.  Intensity and amplitudes of the
noise pulses in the range of interest are fall down to zero after some hours.
A residual part of the type (\emph{b}) pulses is not a noise because of its intensity is proportional to the  intensity of cosmic rays and decreases with moving of the detector on a deeper  underground.
It is known that a part of a photomultiplier own noise pulses could be created
by the charged particles from an outer radioactive background and cosmic rays and also by
charged particles appeared in decays of radioactive residual impurity isotopes in the PMT
construction materials \cite{p6}.
A separate investigation was done for an alignment of a type ($b$) appearing mechanism. The two possibilities were examined.
The first one is a direct generation of the primary electrons from the photocathode or dynode by the cosmic rays.
The second one is an appearing of a photoelectron from the photocathode as a result of absorption of a Cerenkov radiation created by a charged particle in a glass of the entrance window of the PMT.
For the last case, the cosmic rays coming from directions around the vertical can generate in the window a Cerenkov radiation directed outside.
This light will  return into the PMT after the reflection from the bottom sell walls.
The PMT entrance window  was covered by the black paper to check this assumption.
The pulses with a short front only  remained in the spectrum.
Their intensity is changed with the cosmic rays intensity at a moving  of the detector from the second floor of a laboratory building to the ground one.
A conclusion that the short pulses are generated by cosmic rays in the photocathode or dynode has been   made.

A set of measurements was made with the detector at the ground and underground conditions of the BNO INR RAS.
They are:
\begin{enumerate}
  \item Deep Underground Low Background Laboratory (DULB-4900). An unshielded place located at the underground hall under the mountain thickness equal to the 4900 m of the water equivalent (m w.e.). First spectrum (\emph{a}) was measured without any shielding materials and the second one (\emph{b}) was measured with the $0.1\times100\times100$ cm$^3$ Cd sheet (absorber) placed under the detector;
  \item DULB-4900 low background compartment with the walls made of 25 cm polyethylene $+0.1$~cm Cd$+15$~cm Pb. A spectrum (\emph{a}) was measured;
  \item ``2'' + thermal neutron source;
  \item Low background underground laboratory ``KAPRIZ'' at the 1000 m w.e. The spectra (\emph{a}) and (\emph{b}) were measured;
  \item Low background underground laboratory ``NIKA'' at the 660 m w.e. The spectra (\emph{a}) and (\emph{b}) were measured;
  \item Underground hall of the ``CARPET-2'' set-up muon detector at the 5 m w.e. The spectra (\emph{a})
and (b) were measured;
  \item Ground building ``ELLING'' of the ``CARPET-2'' set-up. The spectra (\emph{a}) and (\emph{b}) were
measured;
  \item The open soil. The spectra (\emph{a}) and (\emph{b}) were measured;
  \item The fourth floor (room 404) of the four-storey laboratory building (LAB). The spectrum (\emph{a}) was measured;
  \item The river side of the second floor (room 204) of the four-storey laboratory building. The spectra (a) and (b) were measured;
  \item The valley side of the second floor (room 211) of the four-storey laboratory building. The spectrum (\emph{a}) was measured;
  \item The ground floor of the four-storey laboratory building. The spectrum (\emph{a}) was measured.
\end{enumerate}

The objects are placed in a list in an accordance of a thickness of the cosmic rays absorber above the installation in a sequence of 4900 m w.e.$\rightarrow$ 0 m w.e.$\rightarrow$1.3 m w.e. A brief description of the objects is given in \cite{p11}.

\begin{figure}[pt]
\includegraphics*[width=2.65in,angle=0.]{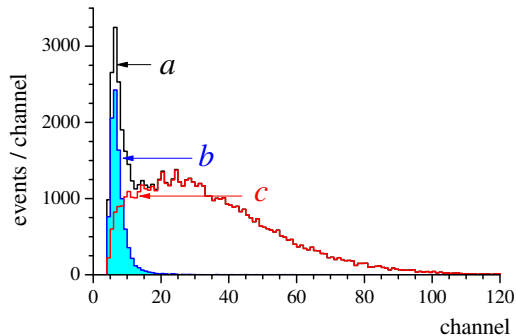}%
\caption{\label{fig:Pulse_amplitude_spectra}
Pulse amplitude spectra of the detector in the point (10) collected at 63.12 h):
``$a$'' -- a total spectrum, ``$b$'' -- a spectrum of pulses with a short front,  ``$c$'' -- a spectrum of pulses with a long front.
}
\end{figure}

\begin{figure}[pt]
\includegraphics*[width=2.5in,angle=0.]{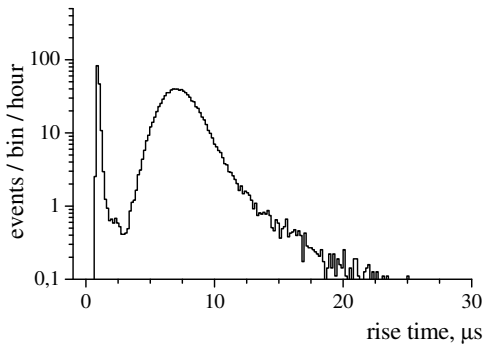}%
\caption{\label{fig:Front_duration}
Front duration distribution of pulses from the spectrum ($a$) from the Fig.~\ref{fig:Pulse_amplitude_spectra} normalized for 1 hour.
}
\end{figure}

A spectrum ``a'' was preliminary measured in the point (10) at
63.12 h to obtain a general imagine about statistical characteristics of pulses. The spectrum is shown
on Fig.~\ref{fig:Pulse_amplitude_spectra},
spectrum ($a$). A distribution of the pulse front durations (0.2-0.8 region of the
total amplitude normalized for 1 hour) is shown on the Fig.~\ref{fig:Front_duration}.
The peaks in the regions of 0.5-1.6 $\mu$s and 3.2-13~$\mu$s are visible.
The last one is correspond to the pulses from the scintillator.
A selection of pulses correspondent to this marked regions allows one to separate the spectrum
($a$) on the Fig.~\ref{fig:Pulse_amplitude_spectra} at fast [spectrum ($b$)] and slow [spectrum ($c$)] components.
It is seen that the fast pulses contribute a main part at the low amplitudes. The more detailed analysis
of the fast pulse shapes shows that the pulses have different decay shapes. Such difference
could be explained by small variable contributions of the scintillation light to the PMT
Cherenkov light. A proportion of the two components depends on a number of particles in an
event and a quantity of tracks crossing the two light generators.  A thin scintillator has a
low sensitivity to the cosmic rays and electrons. Amplitudes of pulses of the events with the
cosmic rays particles crossing of the scintillator are small and lie below the
registration threshold.

Significant number of extraneous pulses could presence in the data sets measured with a low count rate in the underground conditions. Pickups from a periodic switching of industrial equipment and furnishings are the main component of such pulses. They have shapes differ considerably from the useful ones. The extraneous pulses could be excluded from the spectra by using of the discrimination on a base of a pulse shape analysis.

\begin{figure}[pt]
\includegraphics*[width=2.5in,angle=0.]{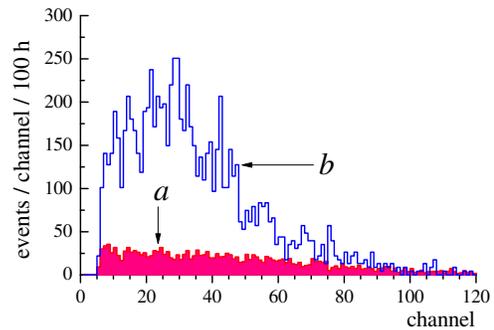}%
\caption{\label{fig:background_pulse_amplitude}
The detector own background pulse amplitude spectrum ``$a$''  and the spectrum of a neutron calibration source ``$b$''.
}
\end{figure}

\begin{figure}[pt]
\includegraphics*[width=2.5in,angle=0.]{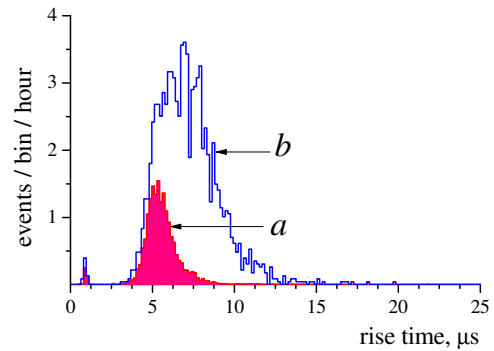}%
\caption{\label{fig:Front_duration_distributions}
Front duration distributions of pulses from the spectrum ($a$) (curve ``$a$'') and spectrum ($b$) (curve ``$b$'') from the Fig.~\ref{fig:background_pulse_amplitude} normalized for 1 hour.
}
\end{figure}

The detector own background should be known for the low count rate measurements.
A measurement at the point (2) was done for this purpose.
The spectrum normalized at 100 h is shown on the Fig.~\ref{fig:background_pulse_amplitude} [spectrum ($a$)].
The corresponding distribution of the front durations is shown on the Fig.~\ref{fig:Front_duration_distributions} (curve ``$a$'').

It is seen from a comparison of the spectra on the Fig.~\ref{fig:background_pulse_amplitude} that the background spectrum has a
longer energy extension than the neutron one. Its front distribution is shifted to the shorter
times simultaneously. A conclusion could be done that the detector background created by
strongly ionized particles with the energy larger than the energy of the products of the
neutron reactions. The $\alpha$-particles from decays of the $^{232}$Th and $^{238}$U natural long lived radioactive isotopes and its daughters contained as micro impurities in the scintillator could be possible sources of the background. The $^{210}$Po ($T_{1/2}=138.4$ d, $E_\alpha=5.3$ MeV) generated in a decay chain of the $^{210}$Pb ($\beta^-$ - decay, $T_{1/2}=21.8$ y) could be another background source.
The last isotope was born charged in the air in the radon decay chain and deposited at the
scintillator charged surfaces during its preparation. This source could be essentially
suppressed by using of the special protecting arrangements against the radon and its
daughters penetration into the gas environment of the scintillator production area. The
$\alpha$-particles which were born outside of the scintillator plate will be absorbed in the covering lavsan film and will not give any noticeable effect.

Estimated path lengths of $\alpha$-particles with the energies of 2051 keV, 4800 keV and 5300 keV in the scintillator with a density of 3.0 g$\cdot$cm$^{-3}$ are equal to 10.6 $\mu$m, 28.6 $\mu$m  and 32.1 $\mu$m \cite{p8}. The one of the 2735 keV triton is 56.3 $\mu$m. The energy spectrum extension will be longer due to the energy difference if the main source of the background is not the neutron reactions products but the $\alpha$-particles from the $^{210}$Po decays. A difference could increase additionally if a specific light yield of the ZnS(Ag) decreased at lower particle energy with an ionization density rise as it occurs for the other nonorganic scintillators \cite{p4}. The light output for the two reaction products with the 4800 keV sum energy will be less than the one for the $\alpha$-particle with the same energy. The lesser front duration of the background pulses could be explained by a location of the $\alpha$-source on a surface of the scintillator grains. Trajectories of $\alpha$-particles will be directed into the grains in this case and the main part of energy will be released inside the ones. It is mentioned in the \cite{p4} that a scintillator based on the dispersed ZnS(Ag) has a long afterglow. It could be explained possibly by a relative increasing of long-lived excitation traps number located on grain surfaces. A particle absorbed in the grain surface layer will give a longer deexcitation time in this case. Vertexes of neutron reactions are distributed uniformly in a $^6$LiF component volume. The reaction products should exit from this materials and fall into the surface layer of the ZnS(Ag) grain to produce a scintillation. An ionization density increases with a particle energy decreasing in accordance with the energy loss dependence \cite{NIST}. A considerable part of the energy will be released in the surface layers of the two adjacent grains giving longer pulse front duration in comparison with a particle absorbed inside the grain.  The features of the detector background distributions mentioned above could be an indirect evidence of a background source grains surface location.

Count rates of the detector at 1 hour for the data integrated above the third channel of the spectra ``\emph{b}'' and ``\emph{c}'' Fig.~\ref{fig:Pulse_amplitude_spectra} are presented in the Table 1 for the all objects cited in the list.

\begin{table*}[ht]
\caption{\label{T1}Count rates of the detector at 1 hour for the data integrated above the third channel of the spectra ($b$) and ($c$) Fig.~\ref{fig:Pulse_amplitude_spectra} and the thermal neutron flux densities.
}

\begin{center}

\begin{tabular} 
{l  c  c  r  c  r  c  r  c}
\hline \hline

\multicolumn{4}{c}{~}         & \multicolumn{3}{c}{Count rate,hour$^{-1}$ } & ~ &    \\

No.& ~& \multicolumn{1}{c}{Place,}& ~& \multicolumn{3}{c}{(3-256 channel)} &~& Thermal neutron flux density,       \\
                                   \cline{5-7}
~& ~& \multicolumn{1}{c}{conditions,}& ~& Short pulse rise time & ~& Long short pulse rise time  &~& (s$^{-1}\cdot$cm$^{-2}$)    \\

\cline{1-1} \cline{3-3}  \cline{5-5}  \cline{7-7} \cline{9-9}

1a & ~& DULB-4900                  & ~& 0.13$\pm$0.03 &~& 21.3$\pm$0.4          &~& (2.6$\pm$0.4)$\times10^{-5}$ \\
1b & ~&  -//- + (Cd)               & ~& 0.09$\pm$0.03 &~& 18.7$\pm$0.5          &~& (1.2$\pm$0.4)$\times10^{-5}$ \\

2  & ~&  DULB-4900 & ~& 0.10$\pm$0.02 &~& \textbf{16.4$\pm$0.3} &~& $\leq3.8\times10^{-7}$ (90\% C.L.)\\
  ~& ~& (low background) & ~& ~ & ~& ~ & ~& ($^3$He prop. counter) \\

3  & ~&  -//- + (n-source)         & ~& 0.2$\pm$0.1   &~& 80$\pm$2              &~& (3.4$\pm$0.4)$\times10^{-4}$ \\

4a & ~&  KAPRIZ                    & ~& 0.13$\pm$0.03 &~&  16.7$\pm$0.3         &~& $\leq5.9\times10^{-6}$ (90\% C.L.) \\
4b & ~&  -//- + (Cd)               & ~& 0.13$\pm$0.03 &~& 16.5$\pm$0.4          &~& $\leq5.9\times10^{-6}$ (90\% C.L.) \\

5a & ~&   NIKA                     & ~& 0.19$\pm$0.04 &~& 17.8$\pm$0.4          &~& (7.5$\pm$3.1)$\times10^{-6}$ \\
5b & ~&  -//- + (Cd)               & ~& 0.14$\pm$0.03 &~& 16.9$\pm$0.3          &~& (3.4$\pm$3.1)$\times10^{-6}$ \\

6a & ~&  $\mu$-detector            & ~& 8.4$\pm$0.5   &~& 64$\pm$1              &~& (2.8$\pm$0.3)$\times10^{-4}$ \\
6b & ~&  -//- + (Cd)               & ~& 8.2$\pm$0.4   &~& 43$\pm$1              &~& (1.4$\pm$0.2)$\times10^{-4}$ \\

7a & ~&  ELLING                    & ~& 23$\pm$1      &~& 1415$\pm$8            &~& (7.5$\pm$0.6)$\times10^{-3}$ \\
7b & ~&  -//- + (Cd)                 & ~& 22$\pm$3      &~& 730$\pm$16            &~& (3.8$\pm$0.4)$\times10^{-3}$ \\

8a & ~&  Open soil                 & ~& 28$\pm$3      &~& 1704$\pm$31           &~& (9.0$\pm$0.8)$\times10^{-3}$ \\
8b & ~&  -//- + (Cd)               & ~& 27$\pm$3      &~& 702$\pm$14            &~& (3.7$\pm$0.3)$\times10^{-3}$ \\

9  & ~&  LAB,  404                 & ~& 22$\pm$1      &~& 1439$\pm$9            &~& (7.6$\pm$0.6)$\times10^{-3}$ \\

10a& ~&  LAB, 204                  & ~& 16$\pm$2      &~& 866$\pm$14            &~& (4.5$\pm$0.4)$\times10^{-3}$ \\
10b& ~&  -//- + (Cd)               & ~& 16$\pm$1      &~& 482$\pm$6             &~& (2.5$\pm$0.2)$\times10^{-3}$ \\

11 & ~&  LAB, 211                  & ~& 19.1$\pm$0.9  &~& 672$\pm$5             &~& (3.5$\pm$0.3)$\times10^{-3}$ \\

12 & ~&  LAB, ground               & ~& 6.5$\pm$0.5   &~& 240$\pm$3             &~& (1.2$\pm$0.1)$\times10^{-3}$ \\

\hline \hline

\end{tabular}
\end{center}
\end{table*}

Values of a thermal neutron flux $F$ in the all examined points could be obtained from the
neutron count rates. A flux $F$ of particles according to a definition is a ratio of a
particles number $\Delta N$ falling on a given surface at a $\Delta t$ time interval to this interval:
$F=\Delta N/\Delta t$. The measured count rate \textbf{n} connected with $F$ by the $ \textbf{n}=\varepsilon \times F$ relation where $\varepsilon$ is a neutron registration efficiency of the scintillation plate. A neutron flux density parameter $\phi$ is used usually to a lightening of a comparison of results obtained with the different geometry detectors. The $\phi$ according to a definition is a ratio of a particles flux $dF_S$ penetrated into the volume of an elementary sphere to the area of it's central cross section $dS: \phi=dF_S/dS$. The $dF_S$ value could be determined using a specific neutron flux falling on an elementary square of the scintillator as  $dF_S=4\times F/S$ where $S$ is an area of the scintillator plane and the coefficient 4 is equal to the ratio of the sphere surface area to the area of the sphere cross section.
The $\phi$ is equal to $\phi=4n/(2S\varepsilon)$ as a result.

It is seems impossible to calculate $\varepsilon$-value because of uncertainties of a composition and a structure of a scintillator layer. This value was obtained experimentally from a  comparison of the $n_1$ count rate of the described detector and $n_{1(2)}$ count rate of the detector with an additional similar passive scintillator plate (2) put under the active plate (1). The plate  (2) was light intercepted. The detector in the measurements was shielded by  the 1 mm cadmium foil for the thermal neutron coming on from the upper hemisphere to shape a  single-sided neutron flux. A count rate of the standard detector is equal to $n_1=\varepsilon_1 \times F$ and the one for the modified detector is $n_{1(2)}=\varepsilon_{1(2)} \times (F-\varepsilon_2 \times F)$ where $\varepsilon_1$ is a plate (1) absorption efficiency of a thermal neutron flux $F$,
$\varepsilon_{1(2)}$ is the absorption efficiency of thermal neutron flux plate (1) in the presence of isolation from the light of the plate (2) absorb some of the flux $F$ with efficiency $\varepsilon_2$.
A value of efficiency is an integral characteristic of a process of absorption for the neutrons coming at different angles and depends on a path passed by a neutron in the scintillator.
An angular distribution of neutrons after passing the one scintillator layer is pulled in the direction normal because of emptive absorption of particles coming at odd angles.
An absorption will be lower for the passed neutrons and $\varepsilon_{1(2)}$ will be lower than $\varepsilon_1$ ($\varepsilon_{1(2)} \leq \varepsilon_1$). The count rates are specified by the expressions $n_2=\varepsilon_2 \times F$ and $n_{2(1)}=\varepsilon_2(1) \times (F-\varepsilon_1 \times F)$ in a case when the plate (2) is used as the active one. Five unknown variables $\varepsilon_1$, $\varepsilon_{1(2)}$, $\varepsilon_2$, $\varepsilon_{2(1)}$ and $F$  are in four obtained equations. One needs to measure additionally a total count rate $n_{[1+2]}=n_{[2+1]}$ for the case  when the both plate used in the active mode to determine precisely all five values. Such measurement is possible if the plates emit scintillation light into the both hemispheres in the detector having two PMTs. The task could be solved with the reviewed detector if the plate adsorbs neutron not strongly.  It could be taken that $\varepsilon_{1(2)}\approx \varepsilon_1$  and $\varepsilon_{2(1)} \approx \varepsilon_2$ in this case. Simple conversions give the expressions  $\varepsilon_1=[n_1-n_{1(2)}]/n_2$ and $\varepsilon_2=[n_1-n_{1(2)}]/n_1$.

The measurements were done in the point (9) of the objects list. The result $\varepsilon_1= \varepsilon_2=0.14 \pm 0.01_{\rm{stat.}} \pm 0.02_{\rm{syst.}} $ was obtained. A difference between $\varepsilon_1$ and  $\varepsilon_{1(2)}$ is calculated for the two homogeneous plate having $\varepsilon=0.14$ at a registration of isotropic neutron flux coming from the one side uses as a systematic uncertainty.

A background count rate measured in the point (2) was subtracted from the data in a process of a determination of a thermal neutron flux density in the each other point. The obtained $\phi$-values are resulted in the last column of the Table \ref{T1}. A limit of a thermal neutron flux density in the point (2) was obtained on a base of measurements with a CH-04 neutron proportional counter with $^3$He \cite{p13}.
\\

\textbf{Discussion of results}

A specific own background of the scintillator plate was found to be to (2.69$\pm$0.05) h$^{-1}\times$(100 cm)$^{-2}$ in the point (2). This value is comparable with a value of a surface $\alpha$-activity of the commercial copper and steel samples which is $\sim(0.5-1.0)$
h$^{-1}\times$(100 cm)$^{-2}$  \cite{p9}.  A surface $\alpha$-activity of the silicon semiconductor samples could reach $\sim 0.1$ h$^{-1}\times$(100 cm)$^{-2}$. It is seems possible to prepare a scintillator plate with similar
surface $\alpha$-activity by using of a specially selected low background [ZnS(Ag)+$^6$LiF] material and a clean technology for the plate preparation. A sensitivity of such scintillator detector (SD) for the thermal neutrons would be comparable with a sensitivity of the $^3$He-proportional counter. A present ratio of the sensitivities is $\sim 16$ as it is seen from a comparison of the data for the (2) and (4) points of the Table \ref{T1}. A difference of the front time duration distribution of background and neutron pulses could be useful for
the plate sensitivity improving also. The obtained results are in a good agreement with the one measured by the $^3$He-proportional counter in the Ref.~\cite{p10}.

A value of the own background defines a sensitivity of the neutron measurement in the
deep underground conditions as it seen from the Table \ref{T1}. A count rate of neutrons in the  DULB-4900 (point (1a)) is equal to (21.32-16.4)/6.11=(0.8$\pm$0.08) h$^{-1}\times$(100 cm)$^{-2}$ with the  effect to background ratio equal to $\sim 0.3$. The neutrons are born in the rock mainly due to  the ($\alpha,n$)-reactions with the light elements. Walls of the ``KAPRIZ'' laboratory are covered with a 30 cm layer of a low background concrete made on a base of a dunite crushed rock.  The concrete is decreased considerably a neutron flux from the rock. (The dunite concrete was used in the construction of the ``NIKA'' laboratory too). A comparison of neutron fluxes in the ``DULB-4900'' and ``KAPRIZ'' measured with CH-04 proportional counter shows that the concrete decreases a neutron flux at $\sim 5.2$ times. One can estimate an expected neutron effect in the ``KAPRIZ'' using of this coefficient and SD count rate in the point (1a) as (0.15$\pm$0.02) h$^{-1}\times$(100 cm)$^{-2}$. A calculation with the data from the Table \ref{T1} gives a value (0.05$\pm$0.08) h$^{-1}\times$(100 cm)$^{-2}$ which is not contradicts to the estimated one. The last value was used to obtain a limit for the neutron flux density in the points (1a and 1b) at 90\%
C.L. as $2 \times (0.05+1.64 \times 0.08)/3600/100/0.17 =5.9 \times 10^{-6}$ cm$^{-2}$c${-1}$.

The own SD background enters a minor deposit into the detector count rate in the measurements at the ground and shallow underground points where reactions of cosmic rays with element nuclei of the environment are the main source of the neutron.

A comparison of the SD count rates with and without a cadmium absorber shows that the absorber decreased a neutron flux at $\sim 1.9$ times under a ceiling and at $\sim 2.4$ times on the open place. Thus, a ratio of neutron fluxes from the soil and from the atmosphere on the open place is equal to $\sim 1.4$.

A SD count rate of the short front pulses in the underground conditions does not depend practically on a value of the external $\gamma$-quanta background level as it seen from the Table~\ref{T1}.  This noise component could be born directly in the PMT by densely ionizing particles. The pulses could appear at the photocathode as a result of direct generation of electrons by $\alpha$-particles of the window surface $\alpha$-activity. A count rate of such pulses connected with the muon intensity and increased proportionally with its growth.
\\

\textbf{Conclusions}

Measurements of some working characteristics of a thermal neutron scintillator detector prepared on a base of a thin 216$\times$304 mm$^2$ plate of a fine-grained [ZnS(Ag)+$^6$LiF] - scintillator made at the BNO INR RAS. An analysis of shapes of charge pulses recorded by means of a digital oscilloscope has showed that background pulses with a short front created by the cosmic rays directly in the photomultiplier are presented among the working pulses in a low energy region of spectra. Measurements of the detector thermal neutron count rate were made at the ground and underground objects of the BNO INR RAS at different shielding for the cosmic rays.
The inherent background of the detector created by $\alpha$-particles from decays of inner radioactive admixture with a surface $\alpha$-activity  at level of (2.69$\pm$0.05)~h$^{-1}\times$(100~cm$^{-2}$)$^{-1}$ was measured.
A ratio of an effect to a background was equal to $\sim 0.3$ at the underground conditions. The neutron pulses have a shorter front than the background ones. It could be used for a discrimination of the background pulses at $\sim 2$ times with an insignificant rejection of the neutron events at the low neutron flux measurements.

\textbf{Acknowledgement}

The work was carried out in part with the financial support of the Federal
Objective Program of the Ministry of Education and Science of the Russian Federation
``Research and Development in the 2007-2013 years on the Priority Directions of the
Scientific and Technological Complex of the Russia'' under the contract No.~16.518.11.7072 and the ``Russian Foundation for Basic Research'' under the grant No.~14-22-03059.

We are thankful to Yu.V.~Stenkin for providing the samples of the scintillator and numerous useful critical comments.


\end{document}